\documentclass{elsart}

\usepackage{epsfig}

\usepackage{amssymb}
\usepackage{amsfonts}
\usepackage{amsmath}
\newcommand{\scrif}{{\mathcal I}^{+}}

\newcommand{\T}{{\mathbb T}}
\newcommand{\C}{{\mathbb C}}
\newcommand{\M}{{\mathbb M}}
\newcommand{\R}{{\mathbb R}}
\newcommand{\rmd}{{\rm d}}
\newcommand{\rmi}{{\rm i}}

\def\const{{\mathrm const}}

\begin{document}

\begin{frontmatter}

\title{Angular momentum of isolated systems\\ in general
relativity\thanksref{inv}}
\thanks[inv]{This paper is based on a talk given at the WCNA 2008 Session
on Differential Geometry and General Relativity.  I thank Professors Krishan
Duggal and Paul Ehrlich for the invitation.}

\author{Adam D. Helfer%
\corauthref{me}}
\corauth[me]{Corresponding author.  Tel. (573) 882-7283, fax (573)
882-1869.}

\address{Department of Mathematics, University of Missouri\\
Columbia, MO 65211, U.S.A.}

\ead{adam@math.missouri.edu}

\begin{abstract}
This article is a conceptual discussion, for non-specialists,  of what appears
to be a satisfactory solution to the problem of treating angular momentum for
isolated radiating systems in general relativity.  The approach is a development
of one suggested by Penrose, based on twistor theory. While in special
relativity angular momentum is a simple tensorial object, in general relativity
it acquires components in other representations of the Lorentz group as well. 
Remarkably, these other components may be identified with the gravitational
radiation.  Thus special-relativistic angular momentum and gravitational
radiation are two parts of one entity, the general-relativistic angular
momentum.
\end{abstract}

\end{frontmatter}

\section{Introduction}
\label{introduction}

General relativity at once destroys the usual foundations for treating energy,
momentum, and angular momentum, and makes the identification of these quantities
especially desirable.  The gravitational field itself --- the varying curvature
of space--time --- is the obstruction to the existence of isometries, and so to
what are usually considered the foundations of the theory of conserved
quantities.  At the same time, the general coordinate freedom, which is
introduced to compensate for this, means that it is only invariantly defined
quantities which are physically meaningful:  energy, momentum and angular
momentum would be the most important of these.

The importance of treating conserved quantities in general relativity, and the 
difficulties in doing so, were recognized by Einstein early
on~\cite{Einstein1916}.  Einstein was
able to overcome these in the restricted case of energy for weak
perturbations of Newtonian theory~\cite{Einstein1918}; 
and the formula he obtained --- the famous quadrupole formula for gravitational
wave-energy --- was deeply influential.

However, the limitations of the quadrupole formula, and other related
approximations (``pseudotensor'' definitions, short-wave formalisms), 
are being increasingly felt,
particularly with the great amount of analytical and numerical
modeling of strongly general-relativistic systems (colliding black
holes, etc.), and also with increasingly accurate treatments of more
modestly general-relativistic systems (binary pulsars, e.g.) in the
past fifteen years or so.  One
needs to move beyond these approximations.

The question of principle is as important as these practical concerns.  What
{\em are} the correct general-relativistic conserved quantities?  (In fact, {\em
are} there such quantities?  Might we just be asking too much?)  We believe that
general relativity is a deeper theory than either Newtonian gravity or special
relativity, indeed that it subsumes them.  If we can find satisfactory
general-relativistic definitions of energy, momentum and angular momentum, we
may expect these isolate something quite fundamental.

While the quadrupole formula itself is so closely allied to Newtonian theory as
to give us little definitive help with these questions, the issues
Einstein took up as he grappled with the problem of conserved quantities remain
the fundamental ones today.  The issues of what boundary conditions to
impose and how to interpret them, the difficulties in formulating a
suitable relativistic concept of a closed system, the problem of defining
suitable reference frames with respect to which the quantities are
defined, remain critical. 

It might be tempting to think that issues of boundary conditions and
closure are analytic  technicalities and that the difficult questions will be
answered by ``figuring out where to put in the epsilons,'' and that with the
right asymptotic conditions the reference frames would be obvious.
This would be
wrong: progress has required, not only substantial physical insight, but subtle
adjustments in what one considers the foundations of the theories to be. 

Definitive progress on aspects of these problems occurred only around 1960, 
with the work of Bondi~\cite{Bondi1960} and Sachs~\cite{Sachs1962}. These
authors made precise the concept of an isolated radiating system, and did so by
finding a suitable generalization of the Sommerfeld radiation conditions ---
that is, boundary conditions.  They were also able to identify  in a subtle way
asymptotic reference frames, and a measure of such a system's energy--momentum,
and show that it had very attractive formal properties.  Since then, the
evidence in favor of the correctness of the Bondi--Sachs definition has
accumulated (and there has been no evidence against it); it is now accepted.

(It is accepted, but it hasn't been used that much yet!  Despite the
considerable activity modeling general-relativistic systems, researchers have
not yet succeeded in using the Bondi--Sachs energy--momentum in most cases. The
main reason for this is that it is hard to construct the Bondi--Sachs asymptotic
regime from the approaches usually used to integrate the equations of motion. 
This remains an important problem.)

The Bondi-Sachs work gave hope that angular momentum might be
accessible by similar techniques.  However, this has turned out not to
be the case.  I will sketch some of this later.  But for right now,
let me say that the problem is associated with the origin-dependence
of angular momentum.  It has been very difficult to identify a
suitable space of origins on which to base a definition.  Using the points in
space--time doesn't work, and, as we shall see, the difficulties are
bound up very deeply with how gravitational radiation qualitatively
affects the structure of space--time.

Another face of the problem is the lack of existence of symmetries ---
something which (as it happened) could be dodged in the treatment of
energy--momentum, but has to be faced squarely for angular momentum. 
It is the gravitational radiation which destroys symmetry in the
asymptotic regime.

As I noted, this lack of symmetries is very serious, because it
means that what we usually consider to be the foundations of the
theory of angular momentum are absent.  What principles will we adopt
to guide our search, then?  What key properties must some candidate
formula have, to be considered angular momentum?

And, supposing we {\em do} succeed in defining angular momentum
general-relativ\-i\-st\-i\-cally, what will the shift in guiding principles
mean for our conception of the quantity more generally?  
If we must alter our views about what
angular momentum {\em is} in order to pass to general relativity, then
presumably what we {\em took} to be angular momentum in Newtonian theory and
in special relativity was rather a special case, and what we took to
be the foundations were (from the general-relativistic view) rather
special properties which hold only in the special case of negligible
gravitational effects.

The aim of this article is to provide a conceptual discussion of what appears to
be a satisfactory solution to the problem of defining angular momentum for
isolated general-relativistic systems.  (In fact, the solution has much stronger
properties than would have been anticipated on the basis of earlier work.)  The
technical details of the construction have appeared elsewhere~\cite{Helfer2007}.

The main conclusions are:

(a) There are two, closely related, criteria which should govern the
search for angular momentum.  The first is that it be possible to
formulate a notion of {\em conservation}.  The second is that it
should be {\em universal}, in that there should be a 
very broad class of
systems for which it is definable, and the angular momenta of the
different systems in this class should be comparable.

The requirement of conservation might sound like a truism --- after
all, one refers to angular momentum as a conserved quantity --- but in
fact the challenge of defining angular momentum has been such that in
many proposals for solving it no very useful notion of conservation exists.

The requirement of universality is a very
strong restriction which goes to the heart of the difficulties.  
It is this which rules out using the points in space--time for the
origins --- there is no
preferred way of identifying the points in one space--time with those
in another.  In fact, the problem of somehow compensating for the loss
of a space of origins is the main one to be faced.

(b) A suitable definition can be given using twistor theory, by
developing ideas of Penrose's~\cite{Penrose1982}.  The appearance of
twistors is in fact natural, because  
in twistor theory, the points of
space--time appear as secondary, derived, objects, and so the problem
of finding origins for angular momentum is recast.

(c) General-relativistic angular momentum, in the presence of
gravitational radiation, has a qualitatively different character than
special-relativistic angular momentum.  It is
represented not simply by a skew two-index tensor field, but contains
other representations of the Lorentz group as well.  Most remarkably,
those ``extra'' contributions turn out to be precisely a standard
measure of the gravitational radiation.  

Thus general-relativistic angular momentum unites the tensorial,
special-rela\-t\-i\-vistic, angular momentum, with gravitational radiation,
which is to be thought of as the essentially general-relativistic
portion.

In section~\ref{angmom}, I will review the appearance of angular momentum in
Newtonian theory and in special relativity.  In section~\ref{sr}, more detail on
special relativity is given, and in section~\ref{tt} the treatment of
special-relativistic angular momentum by twistors is sketched.  Section~\ref{gr}
discusses isolated systems in general relativity, and section~\ref{bms}
how gravitational radiation results in a different asymptotic structure than in
special relativity.  Section~\ref{amgr} explains the twistorial definition of
angular momentum, section~\ref{spcm} explains the sorts of measures of spin and
angular momentum which result from this, and section~\ref{im} discusses some
implications.

This paper is a conceptual, not a technical treatment.  For a technical
treatment, see ref.~\cite{Helfer2007}.  Background reading adequate for
understanding the material there is ref.~\cite{PR1986}.

\section{Angular momentum before general relativity}\label{angmom}

We encounter angular momentum first in Newtonian mechanics, where we
learn that for an object of momentum ${\bf p}$ at position ${\bf r}$
relative to an origin, the (orbital) angular momentum about this
origin is
\begin{equation} 
  {\bf L}={\bf r}\times {\bf p}\, .
\end{equation}
If the object moves in a central force field (central relative, again,
to the same origin), this will be conserved.  More generally, we learn
that the total angular momentum of a closed system will be conserved.

In special relativity, we learn that angular momentum is really a skew
two-index tensor field,
\begin{equation}\label{sram}
 M_{ab}(x)=\left[
\begin{matrix}
0 &ER_x&ER_y&ER_z\\
             -ER_x&0&-L_z&L_y\\ 
             -ER_y&L_z&0&-L_x\\
             -ER_z&-L_y&L_x&0
\end{matrix}
\right]
\end{equation}
where in addition to the ordinary, spatial, angular momentum the
center of energy $(R_x, R_y, R_z)$ (and the energy $E$ itself) appear.
This is a very beautiful object, and the center-of-energy (or more
properly, the moments $ER_x$, $ER_y$, $ER_z$) appear as the conserved
quantities conjugate to the {\em boosts}, that is, the
velocity-changing transformations, just as the spatial angular
momentum is conjugate to the rotations.  (In equation~(\ref{sram}),
the $x$ on the left indicates the point in space--time, whereas on the
right the subscript $x$ stands for one of the coordinates.)

It's also appropriate to touch on angular momentum in quantum
mechanics, for several reasons.  In the first place, it is good to
remember how strongly angular momentum was bound up with the
development of quantum mechanics.  It is also true that part of the
intuition that led to the Bondi--Sachs construction was actually a
product of considerations about quantizing gravity!  An important
motivation for understanding angular momentum in general relativity
remains the hope that this will give us clues about quantum gravity.

In quantum mechanics, we learn that, in the first place, the angular
momentum is a wholly new sort of object, an operator.  We also learn
that there may exist intrinsic angular momentum, or {\em spin}, so
that the total angular momentum operator is
\begin{equation}\label{qmam}
\hat{\bf r} \times \hat{\bf p} +\hat{\bf S}\, ,
\end{equation}
non-relativistically.  (The hats in eq.~(\ref{qmam}) indicate operators.
However, this equation will be the only such case.  The symbol $\hat{\bf r}$
below will indicate a unit vector.)  In the remainder of this paper, however,
quantum theory, and quantum operators, will play no role.

In fact, while the understanding of spin developed in quantum
mechanics, we now realize that the possibility of spin (as intrinsic
angular momentum) is subsumed in angular momentum even at a classical
level.  In modern relativistic parlance, the {\em spin} is the
irreducible, origin-independent, part of the angular momentum (which
can be given by the formula
\begin{equation}
  S_a=(1/2M)\epsilon _{abcd} M^{bc}P^d
\end{equation}
with $M$ the mass, $P^a$ the energy--momentum, and $\epsilon _{abcd}$
the alternating symbol); we think of $M_{ab}$ as determined jointly by
the spin and the center of mass, assuming $P_a$ is known.

\section{Special relativity}\label{sr}

In order to understand general relativity, we should first review a
few elements of special relativity.

Minkowski space $\M =\{ (t,x,y,z)\}\cong\R ^4$ is the space--time of
special relativity; points in it are called {\em events}, and
represent a point in space at an instant in time.  The key structure
on Minkowski space is its {\em metric} $g_{ab}$, in differential form
\begin{equation}
\rmd s^2=\rmd t^2-\rmd x^2-\rmd y^2-\rmd z^2
\end{equation}
(in units where the speed of light is $c=1$).  This convention, with
one plus and three minuses, is the more physically natural one.  The
time experienced by an observer traveling a path $\gamma$ is $\int
_\gamma \rmd s$.  Because essentially all of special relativity can be
derived from arguments about comparison of clocks, the metric is
central.

A vector $v^a$ in Minkowski space may be classified as {\em timelike}, {\em
spacelike} or {\em null}, according to whether
$g_{ab}v^av^b=(v^t)^2-(v^x)^2-(v^y)^2-(v^z)^2$ is positive, negative or zero. 
Timelike vectors then correspond to speeds less than that of light, spacelike
vectors to speeds in excess of light, and (non-zero) null vectors to the speed
of light.

Note that timelike (and for that matter, 
non-zero null) vectors must have $v^t\not= 0$.  Thus the set of these
vectors has two components, those which are {\em future-directed}
(with $v^t>0$), and those which are {\em past-directed} ($v^t<0$).

The isometry group of Minkowski space is called the {\em Poincar\'e
group}.\footnote{In this paper, reflective motions will play no role.
Strictly speaking, we should define the Poincar\'e and Lorentz groups
to consist of orientation- and time-orientation-preserving isometries.}  
Its structure is formally similar to that of a Euclidean
space:  it is a semidirect product of the translations and the
(relativistic version of the rotations, the) Lorentz group:
\begin{equation}
0\to\mbox{Translations}\to\mbox{Poincar\'e}\to \mbox{Lorentz}\to 0\, .
\end{equation}
The fundamental special-relativistic kinematic conserved quantities
are the energy-momentum $P_a$ (conjugate to the translations), and the
angular momentum $M_{ab}(x)$ (conjugate to the Lorentz motions).

Let's pause to note the important properties of $M_{ab}(x)$.  In the
first place, it is origin-dependent.  In the second-place, its
tensorial character means that it is an element of a particular
representation of the Lorentz group.  
The representations we shall be concerned with are labeled by pairs of integers
or half-integers $(s,j)$.  In special relativity, the angular momentum takes
values in the $s=1$, $j=1$ representation.

(A caution about some confusing terminology:  the same representations play
important roles in quantum theory, but the interpretation is somewhat different.
There, the quantities $s$ and $j$ actually become measures of angular momentum.
In this paper, though, they are just labels of representations.)

\section{Twistor theory and angular momentum}\label{tt}

Twistor theory allows an alternative treatment of special relativity,
mathematically equivalent to the usual one, but in which space--time
appears as a secondary, derived, concept.  (This is the main reason why
twistors form a natural candidate for dealing with the difficulties 
of defining angular momentum in general relativity.)

The twistor space appropriate to special relativity is the space $\T =\{
Z^\alpha \}\cong\C ^4$  of spinors of the conformal group of Minkowski space; it
is naturally equipped with a pseudo-Hermitian norm $Z^\alpha{\overline
Z}_\alpha$ of signature $+{}+{}-{}-$.  One can realize $\T$ as a certain space
of spinor fields on $\M$.  I should hasten to say that for this article no
detailed knowledge of spinors is necessary; it is enough to know that a spinor
is a geometric object, something like a vector or a tensor (really like the
square root of a future-directed
null vector, and, unlike vectors, by its definition bound to the
metric structure of the space--time). 

For simplicity, I will only be discussing the geometric interpretation
of the {\em real twistors}, those with $Z^\alpha {\overline
Z}_\alpha=0$, here.  Each such twistor corresponds to a null geodesic
$\gamma$ together with a parallel-transported tangent spinor $\pi
_{A'}$.   I will write $Z^\alpha\leftrightarrow
(\gamma ,\pi _{A'})$ to indicate this correspondence.  Note that the twistor is
rather delocalized from the point of view of space--time, being associated with
a null geodesic rather than a single point.

The angular momentum of a system is coded in a function
$A(Z)=A_{\alpha\beta}Z^\alpha Z^\beta$ on twistor space.  In the case
of real twistors, the correspondence is very direct:
\begin{eqnarray}
A(Z)&=&\mbox{ the component of } M_{ab}(x) \mbox{ determined by the
spinor }\pi _{A'}\nonumber\\
  &&\mbox{ evaluated at any event }x\mbox{ on the
geodesic }\gamma
\end{eqnarray}
if $Z^\alpha\leftrightarrow (\gamma ,\pi _A')$.  
(The choice of point on $\gamma$ is immaterial when the component is
that corresponding to $\pi _{A'}$.)  Explicitly,
\begin{align}\label{spinm}
&A(Z)=\\
&\left[\begin{matrix} \pi _{0'} &\pi _{1'}\end{matrix}\right]
  \left[\begin{matrix} (L_x-ER_x)-\rmi (L_y+ER_x) & -L_z+\rmi ER_z\\
       -L_z+\rmi ER_z& -(L_x+ER_x)+\rmi (-L_y+ER_x)\end{matrix}\right]
  \left[\begin{matrix} \pi _{0'}\\ \pi _{1'}\end{matrix}\right]\, ,\nonumber
\end{align}
where $\pi _{0'}$, $\pi _{1'}$ are the components of $\pi _{A'}$ in a standard
basis~\cite{PR1986}, 
and the center of mass and angular momentum are evaluated at any point
$x$ on $\gamma$.  If we imagine holding the point fixed, but varying the null
geodesics and their tangent spinors through it, we can recover
from~(\ref{spinm}) all components of
the relativistic angular momentum.

\section{General relativity}\label{gr}

In general relativity, a {\em space--time} is a (smooth, connected,
paracompact) manifold ${\mathcal M}$ equipped with a Lorentzian metric
$g_{ab}$.  At any point, then, we may classify the vectors as
timelike, null or spacelike according to the sign of $g_{ab}v^av^b$.
Again, at each point, the timelike and non-zero null vectors divide
into two components.  We assume that it is possible to choose one such
component continuously over the manifold, and that this choice has
been made, so that we know which vectors are {\em future-directed} and
which {\em past-directed}.  We also assume ${\mathcal M}$ is oriented.

How can we make precise the idea of an {\em isolated} system?  In some sense, we
must say what it means to travel far from the system, and say that in that limit
the system becomes ``self-contained.''  Roughly speaking, this should mean
passing to an appropriate asymptotic regime such that all gravitational effects
are localized inside of it.

There are two main approaches to this, based on whether the sense of traveling
far from the system means going in spacelike or null directions.  (There are
also asympotic regimes in timelike directions, but these do not correspond to a
sense of isolation, since in those directions one usually cannot escape from
the influences of matter.)  The asymptotic spacelike regime might seem, based on
non-relativistic experience, most natural, but it does not allow for a direct
treatment of radiation, a phenomenon of central interest; it is also less well
understood mathematically at present.\footnote{A satisfactory definition of
angular momentum in this regime was given by Ashtekar and Hansen~\cite{AH1978}
in the case that the Weyl tensor is ``asymptotically electric''
there, and
Shaw~\cite{Shaw1983} showed
that Penrose's ideas gave this a natural twistorial interpretation.  The
question of just what the physical significance of the restriction to this case
is remains open.  It is possible the ideas outlined in this paper could
help to lift this restriction, since they do elucidate a parallel issue in the
null asymptotic regime.}

If we wish to study an isolated system and and account for the radiation which
it emits, we are led to consider moving away from it at the speed of that
radiation, here the speed of light, that is, in null directions.  
This leads to the

{\bf Definition.}
We say a system is {\em isolated in the sense of Bondi and Sachs} if
it is modeled by a space--time for which we may
identify a set $\scrif$ of {\em escaping wave fronts} with the
following properties:  (a) each escaping wave front is an equivalence
class of asymptotically parallel abreast null geodesics;
(b) the set $\scrif$ has the topology it would
for Minkowski space, and can be joined to ${\mathcal M}$ as a
hypersurface at infinity; (c) the metric has the same leading
asymptotic form (and this holds locally uniformly), 
as one goes out along these geodesics as it would in
Minkowski space.\footnote{This definition is adapted to treat motion
outwards towards the future.  One can time-reverse the concepts to
treat motion inwards from the past and incoming radiation.}

Of course, the foregoing is not phrased mathematically precisely, but it does
capture the main idea.  (The modern technical concept is {\em weak future
asymptotic simplicity}~\cite{PR1986}.)   There are two things worth noting
here.  First, the definition is in part implicit, because one must identify
which wave fronts count as escaping. Second, with this definition we formalize
the idea of {\em modeling} a system by a space--time.  That is, the space--time
${\mathcal M}$ is {\em not} supposed to represent the entire Universe; it is
simply a clean mathematical way of representing an idealized system.  In
practice, we expect many systems to be very well modeled in this way.  The
question of how to account for corrections due to the fact that real systems are
not perfectly isolated is, however, a very important and hard one. There is no
definitive progress on this (neither for energy, momentum nor angular momentum);
this is the problem of {\em quasi-local} kinematics.

\begin{figure}
\begin{center}
\includegraphics*[width=12cm]{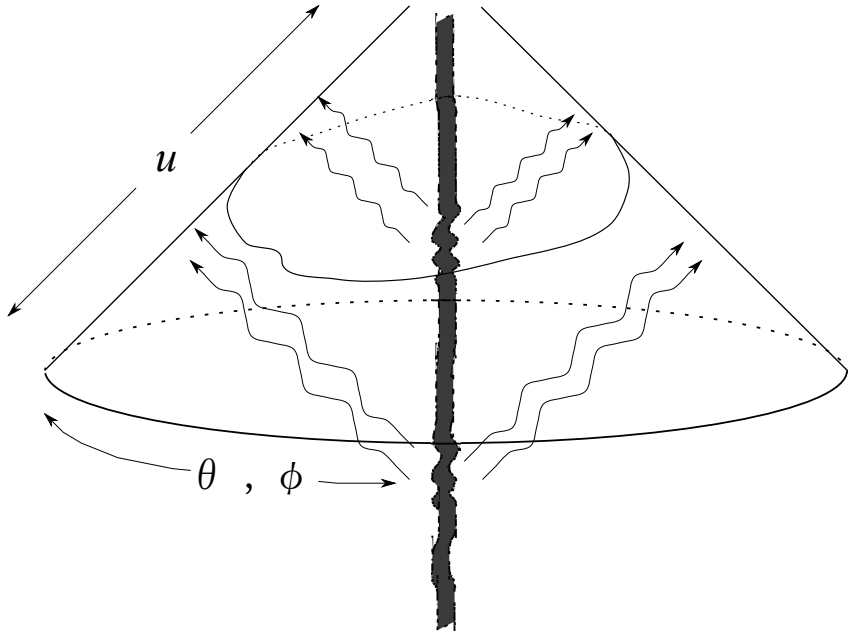}
\end{center}
\caption{An isolated system.
Future null infinity $\scrif$ is the half-cone; the space--time is
below it.  Time
increases upwards, space is horizontal (and one spatial dimension is
suppressed).  The central ``world tube'' represents a region from
which gravitational radiation (indicated
by wavy lines) is emitted; these waves leave their
profiles on $\scrif$.  A generic cut $u+\alpha (\theta ,\phi )
=\const$ is also shown.}
\end{figure}

The hypersurface $\scrif$ at infinity
(called {\em scri-plus}, where ``scri'' comes by
elision from ``script I'') has topology $S^2\times\R$, and this is
easily understood.  The $S^2$ labels the asymptotic angles in which
the wave-fronts might be directed; the $\R$ labels the times (or more
properly, the ``retarded times''  --- what would be $u=t-r$ in
Minkowski space) of emission.

The hypersurface $\scrif$ is evidently the one on which we measure the
outgoing wave profiles of any gravitational radiation; at each point
of $\scrif$ (that is, each outgoing wave front), we give a suitable
measure of the strength of the wave.  It turns out that the strength
of the wave is coded in the asymptotic shear of the outgoing null
geodesics; this is the {\em Bondi shear} $\sigma _{\rm B}(u,\theta
,\phi )$.

In radiation problems, we are interested in measuring the
energy--momentum and angular momentum in a system after some, but
perhaps not all, of the radiation has escaped.  This means that we
seek measures of the energy--momentum and angular momentum in a system
at a retarded time $u$.  However, there is more to it than that.

\section{The Bondi--Metzner--Sachs group}\label{bms}

The group of diffeomorphisms preserving the asymptotic structure of
an isolated system is the {\em Bondi--Metzner--Sachs (BMS) group.}
Its structure is formally very similar to that of the Poincar\'e
group:
\begin{equation}
0\to\mbox{Supertranslations}\to\mbox{BMS}\to\mbox{Lorentz}\to 0
\end{equation}
it is a semidirect product of the Lorentz group (which, recall, is the
relativistic counterpart of the rotations) and the {\em
supertranslations}.
The supertranslations are the motions of the form
\begin{equation}
 u\mapsto u+\alpha (\theta ,\phi )
\end{equation}
(and $\theta\mapsto\theta$, $\phi\mapsto\phi$), where $\alpha$ is any smooth
function.  I will write more about them in a moment.

It is tempting to think that one could use the BMS group as a
``stand-in'' for the Poincar\'e group to define angular momentum.
Approaches based on this idea have not been considered successful,
however.  They lead to mathematical objects with uncomfortably large
functional degrees of freedom and little structure beyond the purely
formal.  The underlying reason for this seems to be that, while the
Poincar\'e group is due to the existence of isometries in Minkowski
space, the BMS group does not represent isometries, but preservation
of a weaker asymptotic structure, which has not been
linked to a physically compelling conserved quantity.

The transition from special to general relativity is marked, here, by
the expansion of the finite-dimensional group of translations to an
infinite-dimensional group of supertranslations.  It is on account of
this difference that the difficulties in defining angular momentum
occur.  The most immediately visible effect is that we no longer have
a finite-dimensional family of preferred measures of time; for we must
consider any retarded time coordinate $u+\alpha (\theta ,\phi )$ just
as good as $u$.  This means too that we have an infinite-dimensional
family of ``instants of retarded time,'' because any {\em cut } of
$\scrif$, of the form $u+\alpha (\theta ,\phi ) =\const$, must be
considered on equal footing with any other (see Fig.~1).  

Why are there supertranslations?  They come about for a direct
physical reason.  Imagine a family of observers very far away (``near
$\scrif$'') from the system in question, located at various angular
coordinates $(\theta ,\phi )$, who synchronize their clocks.  Suppose
a gravitational wave passes, and then suppose they examine their
clocks again.  They will in general find that they have become {\em
desynchronized}, that is, that they have passed from a common retarded
time coordinate $u$ to one of the form $u+\alpha (\theta ,\phi )$.  It
is to accommodate this physical effect that one must introduce
supertranslations.

In special relativity, there are no gravitational waves, this sort of
desynchronization cannot occur, and consequently one has no need to
introduce the BMS group.  But for general relativity, where
gravitational waves are to be expected, we have no choice.

That all cuts $u+\alpha (\theta , \phi )=\const$ of $\scrif$ be on equal footing
has two sorts of consequences for angular momentum.  First, it is bound up with
the problem of the absence of a space of origins about which to measure.  (In
Minkowski space, one can take the $u=\const$ cuts, and their images under
translations, as a preferred set of ``good'' cuts which serve as origins.) 
Second, it affects the sense in which we quantify how much radiation is left in
the space--time after some, but not all, has escaped.  Any cut serves as
a demarcation of ``before'' and ``after'' in this sense.  Thus we seek a measure
of the angular momentum on any cut:  the amount remaining after all radiation
prior to that cut has escaped.\footnote{Thus what will be conserved is the total
angular momentum, comprising that remaining in the space--time and that emitted
in radiation.  One sometimes says that angular momentum at a cut is
``conserved,
but not absolutely conserved.''  While this terminology is a bit odd, 
it has developed in radiation problems, where indeed absolutely conserved
quantities, which are not sensitive to dynamics, are usually not of as much
direct interest.}

I will close this section with a few further comments about the
structure of $\scrif$.  For each fixed $(\theta ,\phi )$, the set of points on
$\scrif$ with differing $u$-values is called a {\em generator} of $\scrif$. 
Then the set of generators is the two-sphere $S^2$, and $\scrif$ naturally has
the structure of a bundle over $S^2$.  This is nothing more than saying that
there is a well-defined space of asymptotic directions for the asymptotic
wave-fronts.

The fact that this space of asymptotic directions is simply $S^2$ plays a deep
role in the analysis.  It turns out that the complex structure on $S^2$ is
determined naturally by the asymptotic geometry.  Because the Lorentz group acts
naturally on this sphere (by fractional linear transformations), one can use
structure to define ``asymptotically constant'' vectors and spinors.  This is a
beautiful and deep feature of the Bondi--Sachs treatment.

\section{Angular momentum in general relativity}\label{amgr}

The approach that I will discuss to defining angular momentum is a
development of ideas of Penrose.  Recall that we are interested in
defining the angular momentum at a {\em cut} $S$ of $\scrif$ (that is,
a set of the form $u+\alpha (\theta ,\phi )=\const$).  Penrose showed
that there was a natural definition of a twistor space $\T (S)$
associated with the cut.  (The elements of $\T (S)$ are spinor fields
on $S$ satisfying certain linear elliptic equations.)  The twistor
space $\T (S)$ is naturally a four-complex-dimensional vector space,
and Penrose found a natural candidate for the angular momentum twistor
$A_S(Z)$ on $\T (S)$.  (He also suggested a definition of the twistor
norm, but pointed out the evidence for this choice was not as strong
as one would like.)

This definition had very attractive formal properties.  The
chief difficulty was that there was no notion of conservation
associated with it.  The angular momenta at two cuts, $S_1$ and $S_2$,
lived in wholly different twistor spaces, $\T (S_1)$ and $\T (S_2)$;
there was no way to begin to compare the angular momentum twistors
$A_{S_1}(Z)$, $A_{S_2}(Z)$.  And physical arguments addressing the
supertranslation problem seemed to provide ``no--go'' theorems, to the
effect that there could be no invariant identification of $\T (S_1)$
and $\T (S_2)$ respecting their natural structures (as complex vector
spaces equipped with certain other natural twistorial data).

However, it turns out that there {\em is} a canonical way of identifying the
twistor spaces.  The trick is that one must be willing to relinquish
(most of) their complex and linear structures, and regard them as real
manifolds.  (The main idea is to show that a spinor field
representing a twistor determines in a preferred way a point on the
cut, and the twistor can then be specified by data at that point.
Those data can then be transported to data at a point on any other cut,
by ``twistor transport'' along the generator of $\scrif$ through the
point. For full details, see~\cite{Helfer2007}.)

Since we have a canonical means of identifying the twistor spaces on
different cuts, we may say that we have a single twistor space
${\mathcal T}$.  This space is intrinsically a real manifold; and
choice $S$ of cut determines a complex-linear structure on ${\mathcal
T}$ which identifies it with $\T (S)$.  Thus we have one twistor space,
with a multiplicity of complex structures.

The space ${\mathcal T}$ also has certain other canonical structures, of which 
one will be important here.   There is a preferred notion of real twistors, and
real twistors are identified with real null geodesics meetings $\scrif$,
equipped with tangent spinors.\footnote{This notion differs from the one derived
from the norm suggested by Penrose for $\T (S)$.}  As I noted earlier, there
{\em is} a well-defined notion of an asymptotic spinor.

As to defining angular momentum on ${\mathcal T}$, we may use
Penrose's definition.  Thus on each cut $S$ we have an angular
momentum $A_S(Z)$.  What is different is that we may now regard all
these angular momenta, for the different cuts, as functions on the
same space ${\mathcal T}$.  We may compute the difference in angular
momenta between two cuts simply as $A_{S_1}(Z)-A_{S_2}(Z)$, and we may
compute the flux of angular momentum by differentiating with respect to
the cut.\footnote{Thus angular momentum is conserved in the sense that we may
now define the angular momentum emitted in gravitational radiation between two
cuts to be the difference in angular momenta on the cuts.  In this sense, the
conservation is rather trivial.  What is not at all trivial, however, 
is the {\em basis} for that statement, the fact that we have a well-defined way
of comparing the angular momenta on arbitrary cuts.}

There is, however, a novel feature, associated with the change in
linear structure as the cut is changed.  Penrose's definition gives
$A_{S_1}(Z)$ as a quadratic form {\em with respect to the linear
structure} $\T (S_1 )$.  Since in general $\T (S_2 )$ will agree with
$\T (S_1)$ as a real manifold but not as a complex vector space, the
function $A_{S_1}(Z)$ would appear as a complicated object, not simply
a quadratic form, on $\T (S_2 )$.  If we wish to compare the angular
momentum at two different cuts, say $A_{S_1}(Z)-A_{S_2}(Z)$, then we
cannot expect this object to appear as a quadratic form on either $\T
(S_1)$ or $\T (S_2)$ (or on any other $\T (S)$).  It will have a more
complicated functional dependence.

This more complicated dependence will appear, when we re-express the
twistorial formulas in more conventional terms, as the angular momentum
not simply taking values in the $s=1$, $j=1$ representation of the Lorentz
group, but having components also in the $s=1$, $j\geq 2$ (for integral $j$)
representations.

\section{Spin and center of mass}\label{spcm}

I mentioned earlier that in special relativity the angular momentum
comprises two parts, the spin and the center of mass.  It is quite
remarkable that the general-relativistic angular momentum can be
decomposed in the same way, with attractive results.

The general-relativistic spin is not simply a vector, but a quantity
which varies over the sphere.  It can be given as
\begin{equation}
\mbox{Spin} (\hat{\bf r} )= {\bf s}_{\rm v}\cdot \hat{\bf r} +M\Im\lambda
(\hat{\bf r} )\, ,
\end{equation}
where $\hat{\bf r}$ is a unit vector representing the direction the
spin is to be measured in, the vectorial part of the spin is ${\bf
s}_{\rm v}$, the mass is $M$, and $\Im\lambda$ is a measure of the
``magnetic'' part of the Bondi shear.\footnote{Just as light waves
have electric and magnetic components, so gravitational waves have what
are called gravitoelectric and gravitomagnetic components.}
That is, the spin can be measured in any direction $\hat{\bf r}$, but these
``components'' ${\rm Spin} (\hat{\bf r})$ 
do not ``integrate up'' to give simply a vector, but
rather a more complicated quantity.
Investigations of angular momentum in general relativity have
repeatedly encountered this magnetic part of the shear, but its role
has been difficult to pin down.  Here we see it is simply the
general-relativistic part of the specific spin.  (``Specific'' meaning
per unit mass.)

While formulas exist for the center of mass, its properties can be
explained more satisfactorily in geometric language.  As the vector
$\hat{\bf r}$ varies over the sphere of directions on the cut in
question, the definition gives, for each $\hat{\bf r}$, a null vector
inwards from the cut which is to be thought of as directed towards the
center of mass of the system.  
The non-linear part of the dependence
of this vector on $\hat{\bf r}$ turns out to be given precisely by
$\Re\lambda (\hat{\bf r})$, a measure of the ``electric'' part of the
Bondi shear.

The picture is actually a bit stronger.  For a system which is asymptotically
stationary, one can use the supertranslation freedom to fix a Bondi retarded
time parameter $u$ for which the electric part of the shear is zero.  The
description of center of mass in the present approach makes the cut at which we
measure look like a ``snapshot'' of such a stationary system, taking into
account that the cut may not be a $u=\const$ one for the retarded time parameter
associated with the stationarity.  This is an appealing physical picture, which
seems to be as strong as one could expect.

What is most remarkable is that, for both the spin and the center of mass, the
essentially general-relativistic portion of the angular momentum is the Bondi
shear, which is a measure of the gravitational radiation.  We may thus say that
general-relativistic angular momentum embraces two parts, a special-relativistic
($j=1$) contribution, and the gravitational radiation ($j\geq 2$).

\section{Implications}\label{im}

We have seen that, in passing from special to general relativity, it
is natural to identify as the angular momentum an object which at once
subsumes the special-relativistic ($j=1$) angular momentum and the
gravitational radiation ($j\geq 2$).  As one might expect, in
situations where general relativity is weak, one gets to good
approximation the conventional, special-relativistic angular momentum
and one can ignore the general-relativistic terms.  But in general one
must include the corrections.

It might be helpful, conceptually, to consider what happens if several
billiard balls collide.  For all practical purposes this is a
Newtonian problem, and one can use non-relativistic angular momentum
to analyze it to excellent accuracy.  However, in principle, when the
balls collide they give off small bursts of gravitational radiation.
These give rise to small supertranslations in the measurements of time
before and after the collisions, and thus to small ambiguities in the
comparison of the Newtonian (or special-relativistic) angular momenta
before and after the collisions.  The general-relativistic definition,
however, resolves these ambiguities. 

For billiard balls, the ambiguities are tiny beyond measurement ($\sim
10^{-53}$ s), but of course they were just to give a homely example.
In (say) the close scattering of two black holes, the resolution of
the ambiguities would be a far more serious matter, necessary to give
any quantitative meaning to angular momentum.

One interesting consequence of this analysis is that it turns out that
angular momentum may be emitted at {\em first order} in the
gravitational wave strength, whereas ordinary energy--momentum is only
radiated at second order.  (Even the $j=1$ part of the angular
momentum can be radiated at first order, although one needs highly
asymmetric and rapidly-changing systems for this to occur.)  This
means that the emission, absorption, or exchange of angular momentum
via gravitational waves may be a much more important feature of
ordinary (not strongly radiating) general relativistic systems, than
the corresponding phenomena for energy--momentum.

While the treatment of angular momentum suggested here does appear to be
satisfactory, in the sense that it is natural and has attractive features, it
leads to deeper questions, which at present we do not have answers for:
Why should the angular momentum have this form?  What underlying structure ---
substituting for the isometries in special relativity --- is responsible for the
existence of angular momentum in general relativity?

\end{document}